\documentclass[aps,reprint,a4paper,nofootinbib]{revtex4-1}
\usepackage[english]{babel}
\usepackage[a4paper,top=2cm,bottom=2cm,left=2cm,right=1.5cm,marginparwidth=1.75cm]{geometry}
\usepackage[utf8]{inputenc}
\usepackage{multirow} 
\usepackage{graphicx,float,lipsum}
\usepackage{xcolor}
\usepackage{amsmath, amssymb}
\usepackage{bbold}
\usepackage[colorlinks=true, allcolors=blue]{hyperref}
\usepackage[font=small, labelfont=bf, margin=5mm]{caption}

\usepackage{natbib}  
      
\begin{document}

\title{\bf A relation between $(2,2m-1)$ minimal strings and the Virasoro minimal string}
\author{Alicia Castro}
\email{blanca-alicia.castro-bermudez@u-bordeaux.fr}
\affiliation{LaBRI,
Université de Bordeaux,\\
351 cours de la Libération, 33405 Talence, France}

\begin{abstract}
  \noindent 
We propose a connection between the newly formulated Virasoro minimal string and the well-established $(2,2m-1)$ minimal string by deriving the string equation of the Virasoro minimal string using the expansion of its density of states in powers of $E^{m+1/2}$. This string equation is expressed as a power series involving double-scaled multicritical matrix models, which are dual to $(2,2m-1)$ minimal strings. This reformulation of Virasoro minimal strings enables us to employ matrix theory tools to compute its $n$-boundary correlators. We analyze the scaling behavior of $n$-boundary correlators and quantum volumes $V^{(b)}_{0,n}(\ell_1,\dots,\ell_n)$ in the JT gravity limit.
\end{abstract}

\maketitle
\medskip

\section{Introduction}

In recent years, a class of 2-dimensional models referred to as dilaton gravity \cite{Grumiller:2002nm} has attracted attention for its rich family of thermodynamic black hole solutions while maintaining the simplicity of gravity in two dimensions. Additionally, it has been shown that a large class of these models are holographically dual to double-scaled matrix models \footnote{See 
\cite{Ginsparg:1993is} for a review on double-scaled matrix models} \cite{Witten:deformationsofJTandmm}.  Within this category, JT gravity \cite{Jackiw:1984je,Teitelboim:1983ux} emerges as a leading example characterized by a linear dilaton potential. While examining JT gravity, another intriguing subfamily of dilaton models emerged, referred to as sinh-dilaton models \cite{Kyono:2017pxs,Turiaci:2020fjj, Mertens:2020hbs}. These models are characterized by a dilaton potential in the form of a hyperbolic sine. A noteworthy instance within this subfamily arises from the investigation of Liouville gravity coupled to a field theory known as time-like Liouville field theory \cite{Gutperle:2003xf}.

Liouville gravity \cite{Polyakov1981QuantumGO,seiberg_notes} is a quantum gravity model that arises from writing the 2-dimensional metric in the conformal gauge, i.e. $g_{ab} = e^{2 b \phi}\hat{g}_{ab}$, where $\hat{g}_{ab}$ is a background/fiducial metric, $0<b<1$ is a constant and $\phi$ is the conformal factor also called the Liouville mode. The action of Liouville gravity is given by $S^L_+$, which can be expressed in terms of the Liouville field theory action
\begin{equation}
    S^L_{\pm}[\phi] = \frac{1}{4\pi}\int d^2z\sqrt{g}\left(\pm g^{ab}\partial_a\phi\partial_b\phi\pm Q_\pm\phi R\pm\mu e^{2b_\pm\phi}\right),\label{eq:Lft_ACTION}
\end{equation}
where $\phi$ is a scalar field, $\hat{g}$ is the fiducial metric, $R$ is its scalar curvature, $\mu_\pm$ is a parameter known as the cosmological constant and $Q_\pm=b\pm \frac{1}{b_\pm}$.

The two sign choices in \eqref{eq:Lft_ACTION} are called spacelike ($+$) and timelike ($-$) Liouville field theory due to the sign of their kinetic term. Each constitutes a conformal field theory (CFT).  
For spacelike Liouville, its central charge is given by\linebreak $c_+=1+6\left(b_++\frac{1}{b_+}\right)^2$, whereas for timelike Liouville, its central charge is $c_-=1-6\left(b_--\frac{1}{b_-}\right)^2$. 

Remarkably, a theory with Lagrangian $S[\phi,\chi]=S^L_+[\phi]+ S^L_-[\chi]$ has been found in the JT gravity literature \cite{Mertens:2020hbs, Suzuki:2021zbe}. This was understood as 2D Liouville gravity coupled to conformal matter in the form of timelike Liouville.
Through field redefinitions and by identifying the metric's conformal factor, the action $S$ can be expressed as a sinh-dilaton gravity action.  Notably, the limit $b\rightarrow 0$ transforms this action into the JT gravity one. Roughly, this is a reflection that in this limit, the dynamics of the conformal factor freeze, and the Liouville gravity path integral reduces purely to the sum over moduli space localized on stable surfaces. 

The key step given in \cite{collier2023virasoro} involves considering a theory with the action $S[\phi,\chi]=S^L_+[\phi]+ S^L_-[\chi]$, treated formally as a Conformal Field Theory (CFT) without associating $\phi$ with the conformal factor of a surface. Instead, mainly CFT machinery is employed\footnote{See related comments on the mathematical robustness of timeline Liouville CFT in \cite{2023_remi}.} to compute its correlation functions, and following the philosophy of JT gravity, the generalization of Weil-Petersson volumes: quantum volumes. 

Within the family of sinh-dilaton models, another noteworthy model is the minimal string, where Liouville gravity couples to a class of CFTs known as Minimal Models \cite{Seiberg:2004at}. These models are labeled by two coprime integers $(p,q)$ characterizing the scaling dimension of primary operators. Expressed in a Lagrangian formulation, minimal models are linked to timelike Liouville through the relation $b=\sqrt{\frac{p}{q}}$ \cite{Zamolodchikov:2005fy}. Specifically, a subset of these models, the $(2,2m-1)$ minimal strings, is known to be dual to $m$-th multicritical one-matrix models \cite{Kazakov:1989bc}. By utilizing the previously mentioned mapping between Liouville field theory and sinh-dilaton gravity, it becomes evident that the limit of minimal strings as $m\rightarrow \infty$ corresponds to JT gravity.

Multicritical one-matrix models serve as the building block for a substantial family of matrix models. The $m$-th multicritical matrix model is characterized by a polynomial matrix potential of order $2m$ and the double-scaled model is characterized by a critical exponent known as the string susceptibility, denoted as $\gamma_s=-\frac{1}{m}$ \cite{Kazakov:1989bc}.

Given the connection between the limit $m\rightarrow \infty$ of $(2,2m-1)$ minimal strings and JT gravity, it is not surprising to find that the theory dual to JT gravity can also be represented as a (infinite) sum of multicritical one-matrix models. This raises the question of whether the Virasoro minimal string can also be understood in terms of a combination of multicritical matrix models.  

In this letter, we propose a relation between the newly formulated Virasoro minimal string and the well-known minimal string. This is accomplished by deriving its genus-$0$ string equation from the density of states obtained in \cite{collier2023virasoro}. The string equation in its power series form makes clear the connection of the Virasoro minimal string with multicritical matrix models and $(2,2m-1)$ minimal strings. This reformulation of the problem enables us to examine Virasoro minimal strings using tools from matrix theory. In particular, we demonstrate how to compute the $n$-boundary amplitudes directly using the string equation and how this gives a valid argument on the small $b$ scaling of the quantum volumes. We also provide commentary on the application of the non-perturbative techniques recently employed for JT (super)gravity \cite{Johnson:2020superJT} and the $b=1$ case.

While preparing this draft, we became aware of related work by Clifford Johnson \cite{Johnson:2024bue}. We are coordinating the submission of our papers.

\section{The Virasoro minimal string}

The Virasoro minimal string, as introduced in \cite{collier2023virasoro}, is a model featuring two CFTs, timelike and spacelike Liouville CFT, described by the action $S[\phi,\chi]=S^L_+[\phi]+ S^L_-[\chi]$. This theory is treated formally as a CFT, with no explicit assignment of $\phi$ to the conformal factor of a surface.
In this framework, the only interaction between timelike and spacelike Liouville CFT is the imposition of the vanishing of the conformal anomaly. This establishes a connection between the central charges of the two theories, given by $c_-=26-c_+$, implying $b_+=b_-=b$.

The claim of \cite{collier2023virasoro} is that the Virasoro minimal string is dual to a doubled-scaled matrix integral with a genus-$0$ density of states

\begin{equation}
  \rho_0(E)=  \frac{2 \sqrt{2} \sinh \left(2 \pi  b \sqrt{E}\right) \sinh \left(\frac{2 \pi  \sqrt{E}}{b}\right)}{\sqrt{E}}.\label{eq:rho0}
\end{equation}

An essential feature of this model is its dependency on the parameter $b$. Notably, as $b$ approaches zero, the leading order of the density of states of the Virasoro minimal string is that of JT gravity. As shown in \cite{collier2023virasoro}, this behavior is also observed in the volumes $V^{(b)}$ and $n$-boundary partition functions, a point we will verify in this letter using the string equation.

\section{String equation}

The string equation is a powerful tool for matrix models and enumeration of maps. In simple terms, it is a transform of the generating function of the number of maps/ribbon graphs.
The full string equation admits a genus expansion and can be written as $u(x)=u_0(x)+\sum_{g=1}^\infty u_g(x)\hbar^{2g}+\dots$, where the dots have been used to refer to non-perturbative contributions that need beyond-genus expansion information \cite{Johnson:2020superJT}. At the planar level, the string equation reduces to $u_0(x)$. For convenience, in this letter, we will refer to the inverse function of $u_0(x)$, $x(u)$, as the string equation.

In general, $x(u)$ is given by a power series 
\begin{equation}
    x(u)=-\sum_{m=1}^\infty t_mu^m,\label{eq:stringeq_times}
\end{equation}
where the coefficients $t_m$ are called times and the value of $u=E_0$ such that $x(E_0)=0$ is called the solution to the string equation. Notably, the string equation assumes a particularly straightforward form when only a finite number of times is non-zero. For instance, in the context of $m$-th multicritical matrix models, it simplifies to $x(u)\sim u^m$. On the other hand, in the case of pure JT gravity \cite{1996CMaPh.181..763K, Mirzakhani}, the times are given by
\begin{equation}
t^{JT}_m=\frac{\pi^{2m-2}}{2\;m!(m-1)!}\label{gamma_k}.
\end{equation}
This implies that JT gravity's dual theory can be expressed as an infinite sum of multicritical matrix models.

Similarly, $(2,2m-1)$ minimal strings exhibit a set of infinitely non-zero times that characterize their string equation \cite{Brezin:1990rb}, which means that minimal strings can also be viewed as an infinite sum of multicritical matrix models. Notably, in the specific scenario of $(2,2m-1)$ minimal strings, their string equation behaves as that of the $m$-th multicritical matrix integral at high energies \cite{Moore:1991ir,Belavin:2008kv}.

A natural question arising is if the Virasoro minimal string admits a string equation description and thus, if its dual theory can be expressed as a sum of multicritical matrix models.  
To investigate this, we assume that $E_0=0$ which allows us to extract the times directly by expanding the density of states \eqref{eq:rho0} in powers of $E^{m+1/2}$, employing a methodology akin to that introduced in \cite{Johnson:2020superJT}.
The line of reasoning is that  if the density of states originates from a double-scaled matrix model described by a string equation $x(u)$, then the disk partition function takes a specific form
\begin{equation}
    Z(\beta)_{g=0}=-\frac{1}{2\sqrt{\pi\beta}}\int_{0}^\infty\mathrm{d} x \frac{\partial x(u)}{\partial u}e^{-u \beta }.\label{eq:Z0_general}
\end{equation}
This expression is derived from matrix models by treating them as quantum mechanical systems with a Hamiltonian featuring the potential $u(x)$. By Laplace transforming this expression, one recovers the original density of states. Therefore, having the density of states allows for reverse engineering this process to deduce the form of the string equation, particularly the times involved.

For the Virasoro minimal string, the expansion of the density of states results in
\begin{equation}
    \rho^{(b)}_0(E)=\sum_{m=0}^{\infty}\sum_{k=0}^{m}\frac{2\sqrt{2}(2\pi)^{2m+2}\left(b^2\right)^{2k-m}}{(2k+1)!(2(m-k)+1)!}E^{m+1/2}.\label{eq:rho_expanded}
\end{equation}
From this expression, we can find the times
\begin{equation}
\begin{aligned}
    t^{(b)}_m=&\sum _{k=0}^{m-1} \frac{\left(2^{5/2}   4^{-m} (2 \pi )^{2 m+1} (2 m-1)!\right) }{(2 k+1)! (m-1)! (2 (m-k)-1)!}\frac{\left(b^2\right)^{2 k-m+1}}{m!}\\
    =& 2 \sqrt{2}\pi\;\frac{ \left(\left(b+\frac{1}{b}\right)^{2m}-\left(b-\frac{1}{b}\right)^{2m}\right) \pi ^{2 m}}{(m!)^2}\label{eq:virasoro_times}.
\end{aligned}
\end{equation}

These times can be considered as a background in topological gravity akin to what is done in \cite{Okuyama:2019xbv}, where each term characterizes the contribution of the $m$-th multicritical matrix model.
By substituting these times into \eqref{eq:stringeq_times}, we obtain the analytical expression
\begin{equation}
    \frac{x^{(b)}(u)}{2\sqrt{2} \pi  }=I_0\left(2 \pi  \sqrt{u}\left(b-\frac{1}{b}\right)\right)-I_0\left(2 \pi  \sqrt{u}\left(b+\frac{1}{b}\right)\right),\label{eq:stringeq_complete}
\end{equation}
where $I_0$ is the modified Bessel function.
The form of this string equation is consistent with the fact that the Virasoro minimal string contains two independent systems (spacelike Liouville with charge $Q_+=b_++1/b_+$ and timelike Liouville with charge $Q_-=b_--1/b_-$) coupled only through a conformal anomaly cancellation which implies $b=b_+=b_-$.

Notably, the Bessel function $I_0$ also appears in JT gravity in the presence of conical defects \cite{Witten:deformationsofJTandmm, Johnson:2020lns}. In this case, the conical defects angles correspond to $\alpha_1=b-\frac{1}{b}$ and $\alpha_2=b+\frac{1}{b}$, with weights $\lambda_1=2\sqrt{2}\pi$ and $\lambda_2=-2\sqrt{2}\pi$. Despite this resemblance, these angles do not conform to typical conical deficits as $\alpha_1<0$ and $\alpha_2>2$ for $0<b<1$. 
Conical defects and other deformations of JT gravity, such as FZZT and end-of-the-world branes, can be viewed as a shift in times, $t^{JT}_m\rightarrow t^{JT}_m + t^{defect}_m$. This is equivalent to having a string equation involving pure JT gravity, defects, or weighted geodesic boundaries \cite{Castro:2023rfd}. However, in the case of the Virasoro minimal string, there is no ‘pure' JT gravity term, and the surfaces differ from hyperbolic surfaces, prompting exploration into the implementation of defects within this framework.

The structure of the string equation \eqref{eq:stringeq_complete} also bears similarity to that of pure $\mathcal{N}=1$ JT supergravity \cite{Johnson:2020superJT}. In the special case where $b=1$, the first term of \eqref{eq:stringeq_complete} becomes constant, and the second term simplifies, precisely aligning with the JT supergravity string equation. We will comment on this in section \ref{section:superjt}.

To validate the consistency of our findings, we compute the disk partition function using the string equation \eqref{eq:stringeq_complete} and the expression
\eqref{eq:Z0_general}. The resulting disk partition function,
\begin{equation}
    Z_0(\beta)=\sqrt{\frac{2\pi }{\beta}} \left(e^{\frac{\pi ^2}{\beta }\left(b+\frac{1}{b}\right)^2}-e^{\frac{\pi ^2 }{\beta }\left(b-\frac{1}{b}\right)^2}\right),
\end{equation}
aligns with \cite[2.17a]{collier2023virasoro}. Similarly, its Laplace transform is consistent with the density of states \eqref{eq:rho0}. The compatibility of these results reinforces the claim that the Virasoro minimal string is dual to a double-scaled random matrix model ensemble. Specifically, this ensemble combines an infinite number of $m$-th multicritical matrix models, akin to the case for JT and JT supergravity.

\begin{figure}
	\centering \includegraphics[width=0.45\textwidth]{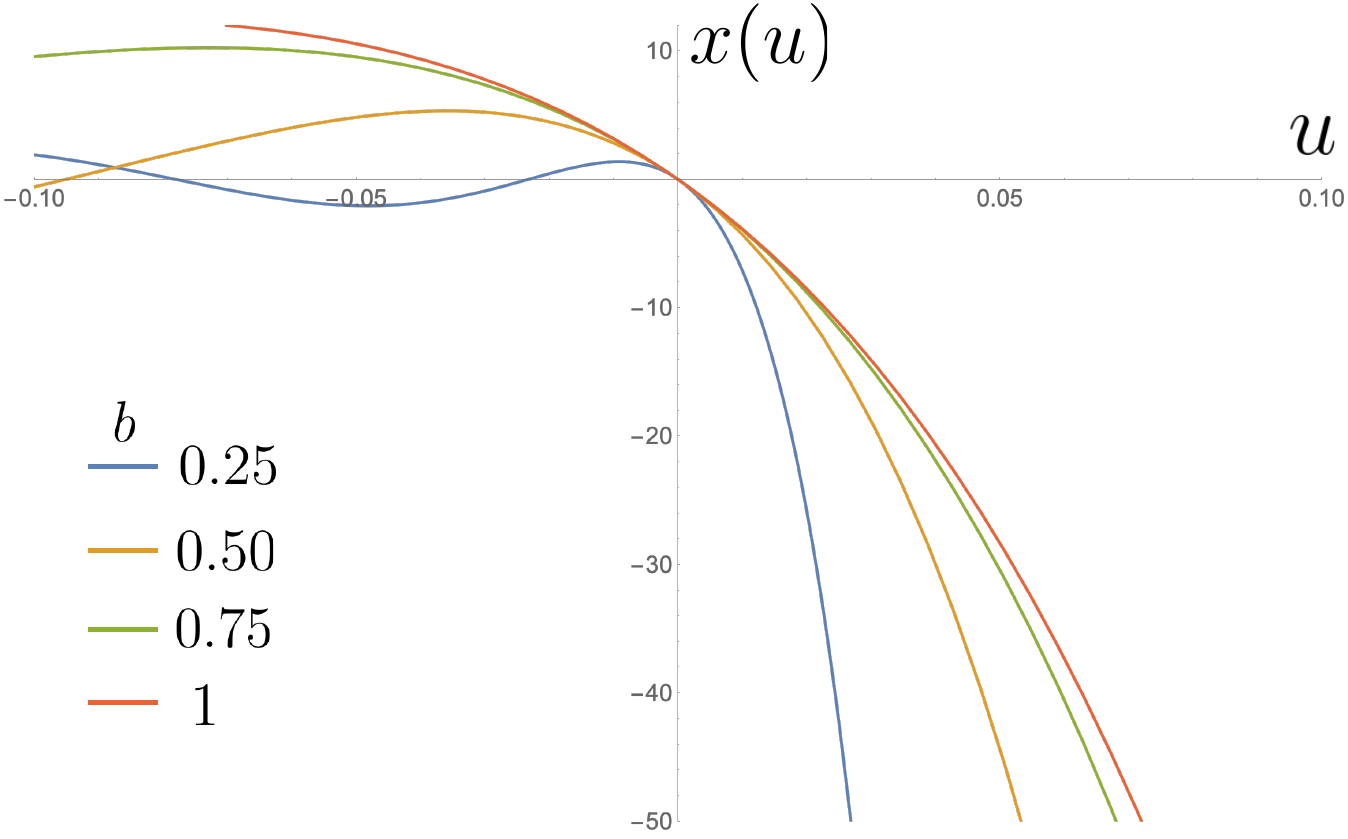}	
	\caption{String equation \eqref{eq:stringeq_complete} for different values of $b$. It is shown that the solution to the string equation is $x(0)=0$ for all $0<b\leq 1$. }
    \label{fig_stringeq_b}
\end{figure}

For completeness, we show a plot of the string equation for various values of $b$ (See Figure \eqref{fig_stringeq_b}).

\section{Spectral curve}
The spectral curve serves as a fundamental element in transitioning from genus zero to a higher genus via topological recursion. It can be obtained by making the change of variables $z^2=-E$ on the genus-$0$ density of states \eqref{eq:rho0}, which results in the double cover of the $E$-plane. This change of variables in \eqref{eq:rho_expanded} results in the spectral curve for the Virasoro minimal string
\begin{equation}
   \frac{y(z)}{2\sqrt{2} \pi}=  \sum _{n=0}^{\infty }\sum _{k=0}^n\hspace{-1mm} \frac{(2 \pi )^{2 n+2} \left((-1)^{n+1} \left(b^2\right)^{2 k-n}\right)}{(2 k+1)! (2 (n-k)+1)!}z^{2 n+1} \label{eq:spectral_density}
\end{equation}
which agrees with the result found in \cite[5.15]{collier2023virasoro}.

In \cite{collier2023virasoro}  the authors employ moduli space analysis tools to determine expressions for the spectral curve in terms of integrals over the moduli space of curves $\bar{\mathcal{M}}_{g,n}$. This expression (\cite[5.21]{collier2023virasoro})  involves the times $\tilde{t}_1=\frac{c-13}{24}(2\pi)^2$ and $\tilde{t}_{2m}=-\frac{B_{2m}(2\pi)^{4m}}{(2m)!(2m)!}$ for $m\geq 1$, where $B_{2m}$ are the Bernoulli numbers. We want to comment on how these times $\tilde{t}$ are related to the times of the string equation \eqref{eq:virasoro_times}. 

The requirement for a match between the series expansion of the spectral curve and the intersection number expansion, as shown in \cite[5.15]{collier2023virasoro}, is expressed by the equation
\begin{equation}
    \sum_{m\geq 0} \hat{t}_m u^m = \exp \left(-\sum_{m\geq 0} \tilde{t}_m u^m\right)\label{eq:modulitimes}.
\end{equation}
Upon a direct analysis, it becomes evident that the times $\hat{t}_m$ are related to the times in this paper ($t_m$) through the relation
\begin{equation}
    \hat{t}_m=2(-1)^{m+1}(m+1)!t_{m+1}.
\end{equation}
Then, by solving \eqref{eq:modulitimes} for $\tilde{t}_m$ using \eqref{eq:virasoro_times}, one finds the result in \cite[5.24]{collier2023virasoro} mentioned above.

\section{$n$-boundary amplitudes and quantum volumes}

Given the evidence supporting the correspondence of the Virasoro minimal string with a double-scaled matrix integral and multicritical matrix models, we employ tools from the matrix model literature \cite{Ambjorn:1990ji,Moore:1991ir} to derive genus zero $n$-loop correlators solely using the string equation. In \cite{Mertens:2020hbs}, it was highlighted that for JT gravity and minimal strings, one can obtain the $n$-boundary correlators using 

\begin{equation}
\begin{aligned}
  \left\langle\prod_{i=3}^n Z_0(\beta_i) \right\rangle= &-\frac{\sqrt{\beta_1\dots \beta_n}}{2\pi^{n/2}} \label{eq:Zn_beta}\\
  & \hspace{-2mm}\left(\frac{\partial}{\partial x}\right)^{n-3}
  \left(u'(x)\; e^{-u(x)(\beta_1+\dots+\beta_n)}\right)_{u\rightarrow 0}.
\end{aligned}
\end{equation}
As mentioned in \cite{Mertens:2020hbs}, we need expressions for the derivatives of $u(x)$; however, our analytic expression for the string equation \eqref{eq:stringeq_complete} is given in terms of $x(u)$. This issue is solved by applying Lagrange inversion to obtain
\begin{equation}
    \left(\frac{\partial^{n} u}{\partial x^{n}}\right)_{u\rightarrow 0}=\underset{u\to 0}{\text{lim}}\frac{\partial ^{n-1}\left(\frac{u}{x(u)}\right)^n}{\partial u^{n-1}}.
\end{equation}
Using this formula and the string equation \eqref{eq:stringeq_complete}, we can derive expressions for the genus-$0$ $n$-boundary amplitudes of the Virasoro minimal string. For instance,
\begin{align}
    &\left\langle \prod_{i=1}^3 \frac{Z_0^{(b)}(\beta_i)}{\sqrt{\beta_i}}\right\rangle=\frac{1}{2 \sqrt{2}\pi (8\pi ^{7/2})}\\
    &\left\langle \prod_{i=1}^4 \frac{Z_0^{(b)}(\beta_i)}{\sqrt{\beta_i}} \right\rangle=\frac{1}{ (2\pi)^8} \left(\pi ^2 \left(b^2+\frac{1}{b^2}\right)+\sum_i^4\beta_i\right)
\end{align}

\begin{equation}
\begin{aligned}
&\left \langle \prod_{i=1}^5 \frac{Z_0^{(b)}(\beta_i)}{\sqrt{\beta_i}} \right\rangle=\frac{1}{2^{12}\sqrt{2} \pi ^{19/2}}\left(5\pi ^4 \left( b^4+\frac{1}{b^4} +\frac{26}{15}\right)\right.\\
    &\hspace{15mm}\left.+6 \pi^2 \left(b^2+\frac{1}{b^2}\right) \left(\sum_{i=1}^5\beta_i\right)+2 \left(\sum_{i=1}^5\beta_i\right)^2\right).
\end{aligned}
\end{equation}
We can go one step further by assuming that, in a similar way as for JT gravity, the $n$-boundary correlators \eqref{eq:Zn_beta} can be decomposed into the volume of surfaces with $n$ geodesic boundaries $V^{(b)}_{0,n}(\ell_1,\dots,\ell_n)$ and $n$ trumpets connecting the bulk with $n$ copies of the asymptotic boundary by gluing them along each geodesic boundary. This is 
\begin{equation}
\begin{aligned}
    \left\langle \prod_{i=1}^n Z_0^{(b)}(\beta_i)\right\rangle =\int_0^\infty  \prod_{i=1}^n & 2\ell_j \mathrm{d} \ell_i \;Z^{(b)}_{tr}(\beta_i,\ell_i)\\
    &\times V^{(b)}_{0,n}(\ell_1,\dots,\ell_n),\label{eq:Zn_virasoro}
\end{aligned}
\end{equation}
where $Z^{(b)}_{tr}(\beta,\ell)$ is the trumpet of the Virasoro minimal string. It is given by

\begin{equation}
    Z^{(b)}_{tr}(\beta,\ell)=\sqrt{\frac{2\pi}{\beta}}e^{-\frac{4\pi^2 \ell^2}{\beta}}
\end{equation}
where $\ell$ can be interpreted as the analogous of the length of geodesic curves dividing the surface in bulk and trumpets.   

In this way, the $n$-boundary correlator \eqref{eq:Zn_beta} provides a generating function for integral transforms of the quantum volumes $V^{(b)}_{0,n}(\ell_1,\dots,\ell_n)$ solely using the string equation. This situation resembles the JT gravity case where the string equation with times \eqref{gamma_k} serves as a building block to construct the generating function of Weil-Petersson volumes \cite{Budd:2020rbm}.

Using this expression for the $n$-boundary correlators of the Virasoro minimal string, we can examine their $b\rightarrow 0$. Notably, to obtain a consistent scaling of the trumpet in the JT gravity limit, one needs to rescale the asymptotic boundary length $\beta=\frac{1}{b^2}\beta^{JT}$ and the `geodesic' length $\ell=\frac{\ell^{JT}}{4\pi b}$ and take the limit $b\rightarrow 0$. It is observed that the product of $n$ asymptotic lengths in the first line of \eqref{eq:Zn_beta} introduces a factor of $b^{-n}$, whereas the derivatives of the string equation in the second line of \eqref{eq:Zn_beta} introduce a leading-order scaling $(b^2)^{3-n}$. Combining these factors, we deduce that
\begin{equation}
    Z^{(b)}_{0,n}(\mathrm{\beta})\rightarrow (8\pi^2 b^2)^{\frac{3}{2}(2-n)}  Z^{(JT)}_{0,n}(\mathrm{\beta}_{JT})
\end{equation}
where $Z^{JT}$ is the genus-$0$ $n$-boundary correlator given by \cite[127]{Saad:2019lba}. By using \eqref{eq:Zn_virasoro} in conjunction with the expression for the Virasoro minimal string trumpet, we can deduce that the quantum volumes scale as
\begin{equation}
     V^{(b)}_{0,n}(\ell_1,\dots,\ell_n)\rightarrow  (8\pi^2 b^2)^{3-n} V^{WP}_{0,n}(\ell^{JT}_1,\dots,\ell^{JT}_n) 
\end{equation}
when $b\rightarrow 0$.
This result aligns with the result obtained in \cite[2.25]{collier2023virasoro}, verifying the explicit dependence on $b$ of the volumes and providing further evidence that the generating function of these volumes emanates from a double-scaled matrix model.

\section{A comment on non-perturbative effects}

The reformulation of the Virasoro minimal string in terms of multicritical matrix models facilitates a non-perturbative analysis originated from matrix models and applied to JT (super) gravity and its deformations in \cite{Johnson:2020superJT}. In this setting, it is noted that the complete non-perturbative string equation is obtained by replacing the formal variable $u^m$ with the Gel'fand-Dikii polynomial $R_m[u]$

\begin{equation}
    \mathcal{R}=\sum_{m=1}^\infty t_mR_m[u]+x.
\end{equation}
Then, the string equation $\mathcal{R}=0$ takes the form \footnote{In fact, a more general equation satisfied by $\mathcal{R}$ is discussed in \cite{Johnson:2020superJT}, for those readers interested in further details.}
\begin{equation}
    x=-\sum_{m=1}^\infty t_mR_m[u]\label{eq:string_eq_nonpertu}.
\end{equation}
To solve \eqref{eq:string_eq_nonpertu}, one typically employs a truncation in $m$. For example, with a truncation up to $m_{\text{max}}=7$ was shown to be enough in the case of JT gravity \cite{Johnson:2020exp}. The value of $m_{\text{max}}$ depends on how fast the times $t_m$ decrease with respect to $m$.

\begin{figure}
	\centering \includegraphics[width=0.45\textwidth]{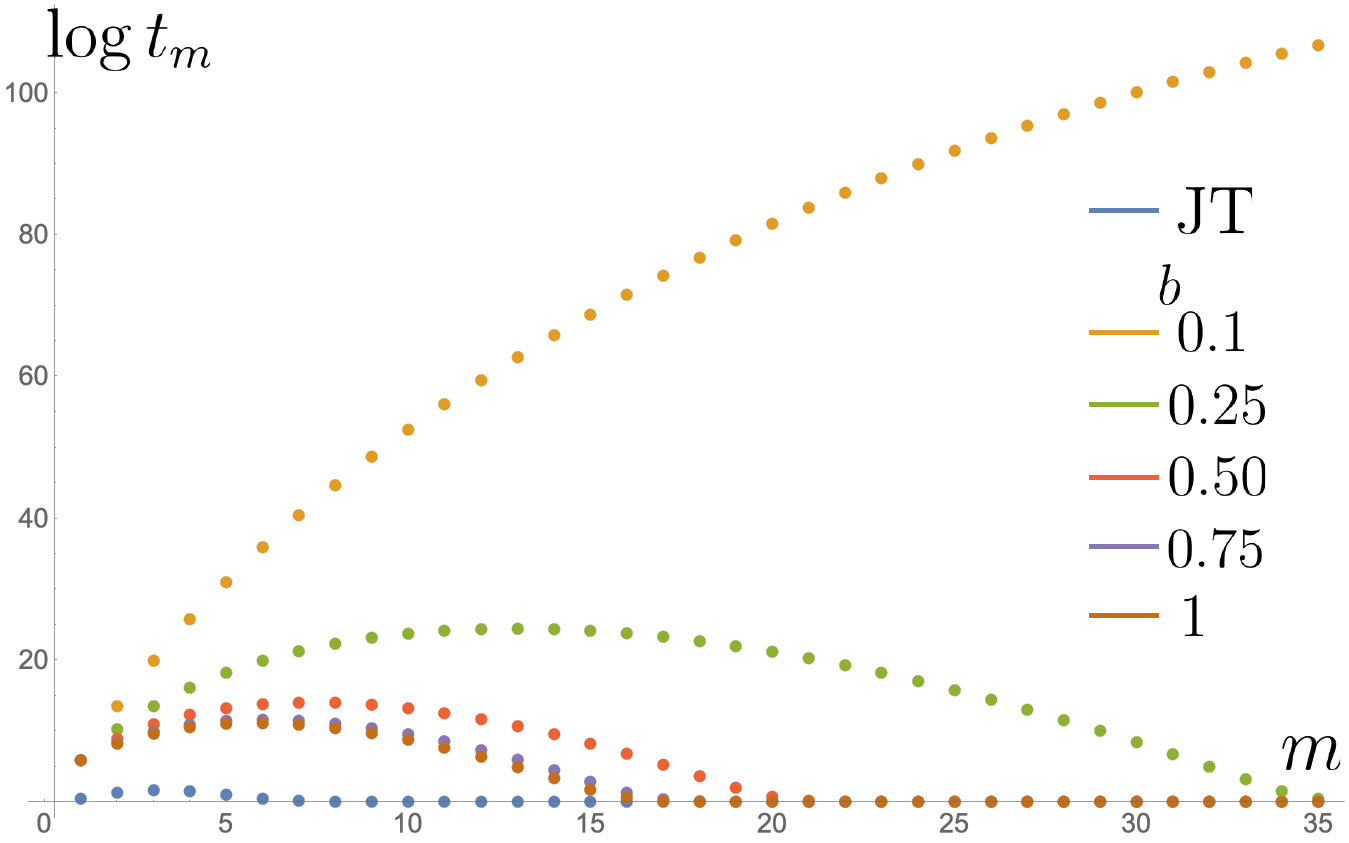}	
	\caption{Logaritmic behavior of the times $t_m$ in \eqref{eq:stringeq_complete}. As $b\rightarrow 0$, the slower they decrease with respect to $m$.}
	\label{fig_times}
\end{figure}

For the Virasoro minimal string, the times \eqref{eq:virasoro_times} follow the behavior $t_m\sim b^{-2m}\left(b^2\right)^{2m-k}$ with $0\leq k\leq 2m$.  Therefore, as $b\rightarrow 0$, an increasing number of times becomes relevant for the non-perturbative analysis (See Figure \ref{fig_times}). This behavior contrasts with the one in JT gravity (blue graph in Figure \ref{fig_times}). 

On the opposite side, the Virasoro minimal string times decrease faster with $m$ as $b\rightarrow 1$. This suggests that achieving the non-perturbative completion of the Virasoro minimal string at $b=1$ might be more feasible than for the JT limit, at least using this tool.

\begin{figure}
	\centering \includegraphics[width=0.45\textwidth]{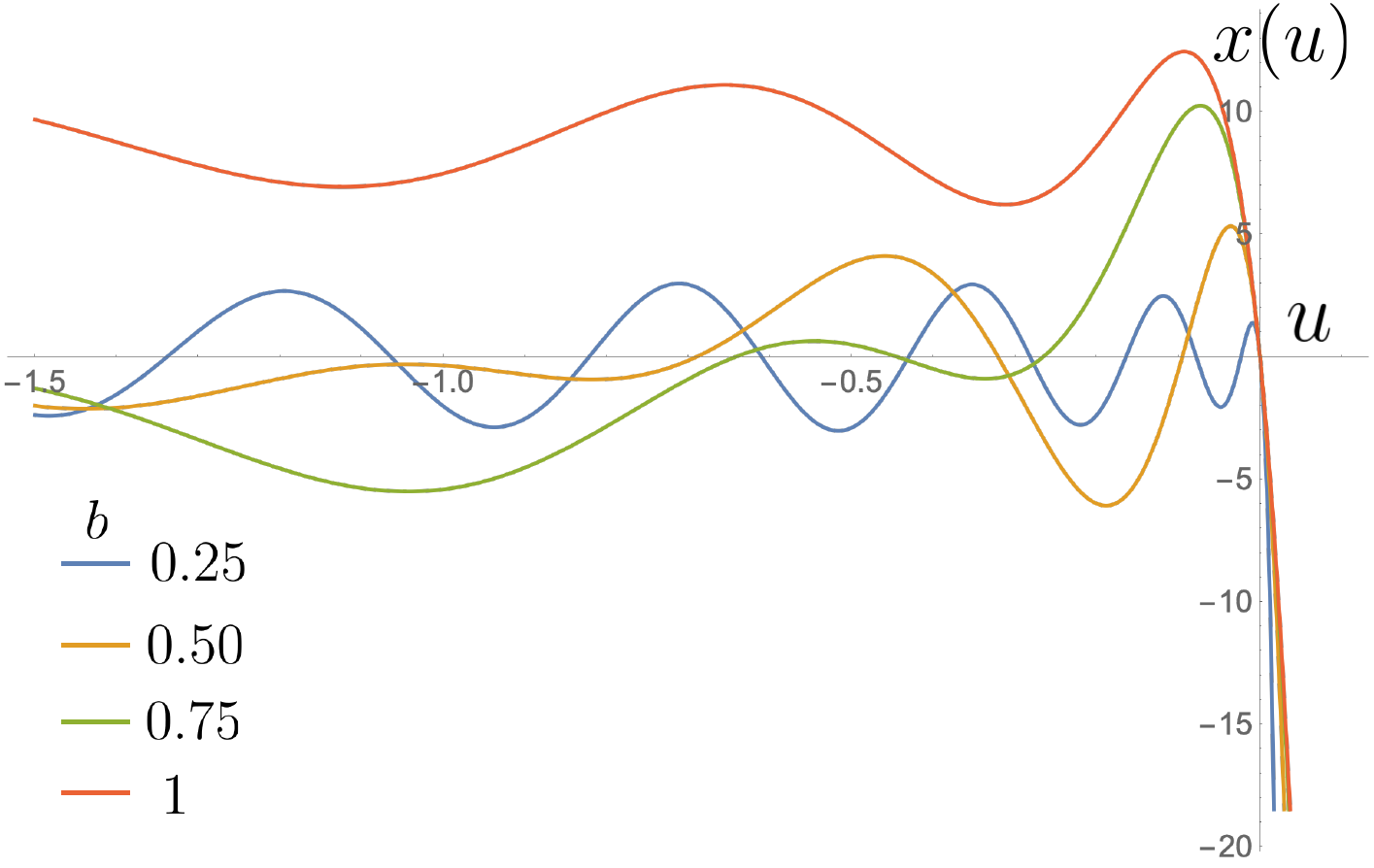}	
	\caption{String equation \eqref{eq:stringeq_complete} for different values of $b$ in the regime $u<0$.}
	\label{fig_stringeq_asymp}%
\end{figure}

It is worth mentioning that a preliminary non-perturbative correction can be derived by ensuring that the string function remains single-valued across the entire real line $-\infty < x < \infty$ \cite{Johnson:2020lns}. Examining the behavior of \eqref{eq:stringeq_complete} for $x>0$, we observe that the value $E_0$ for $0<b<1$ is (at least) close to 0. Conversely, for $b=1$, $x(-\infty) \rightarrow E_0 \neq 0$ (See Figure \ref{fig_stringeq_asymp}). Such analysis was studied in \cite{Johnson:2024bue}.

\section{Relation to $\mathcal{N}=1$ JT supergravity}\label{section:superjt}
In the case $b=1$,
the string equation \eqref{eq:stringeq_complete} takes the form

\begin{equation}
   \frac{x(u)}{2 \sqrt{2} \pi }=  \left(1-I_0\left(2 \pi  \sqrt{4u}\right)\right).
\end{equation}

This expression bears a striking resemblance to the string equation of JT supergravity \cite{Johnson:2020superJT} for $\tilde{u}=4 u$.

While the origin of this connection remains unclear, it may be associated with an observation from \cite{collier2023virasoro} that the matrix model appears to exhibit non-perturbative instability, except when $b=1$. This aligns with the established understanding that the matrix model for super JT gravity exhibits a better non-perturbative behavior compared to its non-supersymmetric counterpart.

This similarity suggests a relation between the dual matrix models of the Virasoro Minimal String and JT supergravity. Moreover, extending the analogy to the level of the worldsheet theories, it becomes intriguing to investigate whether this connection translates into relations between the volumes of moduli and super-moduli space and their integrable structure.

\section{Discussion}
In this letter, we provide a relation between the Virasoro Minimal String and multicritical matrix models and $(2,2m-1)$ minimal strings within the framework of double-scaled matrix models. The key step was to derive the genus-$0$ string equation for the Virasoro minimal string. This allowed us to use tools from random matrix theory to compute the genus-$0$ disk partition function and $n$-boundary amplitudes directly from the string equation. Additionally, it facilitates the study of the Virasoro minimal string in the framework of topological gravity similarly to \cite{Okuyama:2019xbv}. 

In particular, the expression of the $n$-boundary amplitude provides a generating function for integral transforms of the quantum volumes $V^{(b)}(\ell_1,\dots,\ell_n)$. This formulation may allow us to study the Virasoro minimal string as a model of random geometry, up to the caveats remarked on \cite{collier2023virasoro}, by writing the generating function of $V^{(b)}(\ell_1,\dots,\ell_n)$ directly.

Our results for the regime $b\rightarrow 0$ are compatible with those in \cite{collier2023virasoro} even though the starting point is different. This provides further evidence that the dual theory of the Virasoro minimal string is in fact a  double-scaled matrix model. 

Our investigation also found parallels between the Virasoro Minimal String and JT supergravity, particularly in their string equations. The observed behavior as $b\rightarrow 1$ suggests non-perturbative completion possibilities, echoing recent insights into the stability of matrix models. It also highlights the potential to investigate whether this connection manifests as a relation between the volumes of moduli and super-moduli space or their integrable structure.

Some natural following steps would be to analyze further the $b=1$ model and its relation to super JT gravity. In particular, it would be interesting to investigate how the Virasoro minimal string's dual matrix model interpolates between the duals of JT gravity and supergravity. Furthermore, a question left open in \cite{collier2023virasoro} is how to implement defects in the Virasoro minimal string. In this regard, the string equation provides a path to follow similar to that of JT gravity with general defects \cite{Castro:2023rfd} where defects are introduced as extra terms in the string equation.

\section*{Acknowledgements}
I want to thank Clifford Johnson for discussions on minimal models and non-perturbative effects in matrix models and Timothy Budd for useful discussions on random geometry tools for JT gravity. This work was supported by the ANR- 20-CE48-0018 “3DMaps” grant.

\bibliographystyle{elsarticle-num-names} 
\bibliography{bibliography}

\begin{thebibliography}{31}
\expandafter\ifx\csname natexlab\endcsname\relax\def\natexlab#1{#1}\fi
\providecommand{\url}[1]{\texttt{#1}}
\providecommand{\href}[2]{#2}
\providecommand{\path}[1]{#1}
\providecommand{\DOIprefix}{doi:}
\providecommand{\ArXivprefix}{arXiv:}
\providecommand{\URLprefix}{URL: }
\providecommand{\Pubmedprefix}{pmid:}
\providecommand{\doi}[1]{\href{http://dx.doi.org/#1}{\path{#1}}}
\providecommand{\Pubmed}[1]{\href{pmid:#1}{\path{#1}}}
\providecommand{\bibinfo}[2]{#2}
\ifx\xfnm\relax \def\xfnm[#1]{\unskip,\space#1}\fi
\bibitem[{Grumiller et~al.(2002)Grumiller, Kummer, and Vassilevich}]{Grumiller:2002nm}
\bibinfo{author}{D.~Grumiller}, \bibinfo{author}{W.~Kummer}, \bibinfo{author}{D.~V. Vassilevich},
\newblock \bibinfo{title}{{Dilaton gravity in two-dimensions}},
\newblock \bibinfo{journal}{Phys. Rept.} \bibinfo{volume}{369} (\bibinfo{year}{2002}) \bibinfo{pages}{327--430}. \DOIprefix\doi{10.1016/S0370-1573(02)00267-3}. \href{http://arxiv.org/abs/hep-th/0204253}{{\tt arXiv:hep-th/0204253}}.
\bibitem[{Ginsparg and Moore(1993)}]{Ginsparg:1993is}
\bibinfo{author}{P.~H. Ginsparg}, \bibinfo{author}{G.~W. Moore},
\newblock \bibinfo{title}{{Lectures on 2-D gravity and 2-D string theory}},
\newblock in: \bibinfo{booktitle}{{Theoretical Advanced Study Institute (TASI 92): From Black Holes and Strings to Particles}}, \bibinfo{year}{1993}, pp. \bibinfo{pages}{277--469}. \href{http://arxiv.org/abs/hep-th/9304011}{{\tt arXiv:hep-th/9304011}}.
\bibitem[{Witten(2020)}]{Witten:deformationsofJTandmm}
\bibinfo{author}{E.~Witten},
\newblock \bibinfo{title}{{Matrix Models and Deformations of JT Gravity}},
\newblock \bibinfo{journal}{Proc. Roy. Soc. Lond. A} \bibinfo{volume}{476} (\bibinfo{year}{2020}) \bibinfo{pages}{20200582}. \DOIprefix\doi{10.1098/rspa.2020.0582}. \href{http://arxiv.org/abs/2006.13414}{{\tt arXiv:2006.13414}}.
\bibitem[{Jackiw(1985)}]{Jackiw:1984je}
\bibinfo{author}{R.~Jackiw},
\newblock \bibinfo{title}{{Lower Dimensional Gravity}},
\newblock \bibinfo{journal}{Nucl. Phys. B} \bibinfo{volume}{252} (\bibinfo{year}{1985}) \bibinfo{pages}{343--356}. \DOIprefix\doi{10.1016/0550-3213(85)90448-1}.
\bibitem[{Teitelboim(1983)}]{Teitelboim:1983ux}
\bibinfo{author}{C.~Teitelboim},
\newblock \bibinfo{title}{{Gravitation and Hamiltonian Structure in Two Space-Time Dimensions}},
\newblock \bibinfo{journal}{Phys. Lett. B} \bibinfo{volume}{126} (\bibinfo{year}{1983}) \bibinfo{pages}{41--45}. \DOIprefix\doi{10.1016/0370-2693(83)90012-6}.
\bibitem[{Kyono et~al.(2017)Kyono, Okumura, and Yoshida}]{Kyono:2017pxs}
\bibinfo{author}{H.~Kyono}, \bibinfo{author}{S.~Okumura}, \bibinfo{author}{K.~Yoshida},
\newblock \bibinfo{title}{{Comments on 2D dilaton gravity system with a hyperbolic dilaton potential}},
\newblock \bibinfo{journal}{Nucl. Phys. B} \bibinfo{volume}{923} (\bibinfo{year}{2017}) \bibinfo{pages}{126--143}. \DOIprefix\doi{10.1016/j.nuclphysb.2017.07.013}. \href{http://arxiv.org/abs/1704.07410}{{\tt arXiv:1704.07410}}.
\bibitem[{Turiaci et~al.(2021)Turiaci, Usatyuk, and Weng}]{Turiaci:2020fjj}
\bibinfo{author}{G.~J. Turiaci}, \bibinfo{author}{M.~Usatyuk}, \bibinfo{author}{W.~W. Weng},
\newblock \bibinfo{title}{{2D dilaton-gravity, deformations of the minimal string, and matrix models}},
\newblock \bibinfo{journal}{Class. Quant. Grav.} \bibinfo{volume}{38} (\bibinfo{year}{2021}) \bibinfo{pages}{204001}. \DOIprefix\doi{10.1088/1361-6382/ac25df}. \href{http://arxiv.org/abs/2011.06038}{{\tt arXiv:2011.06038}}.
\bibitem[{Mertens and Turiaci(2021)}]{Mertens:2020hbs}
\bibinfo{author}{T.~G. Mertens}, \bibinfo{author}{G.~J. Turiaci},
\newblock \bibinfo{title}{{Liouville quantum gravity -- holography, JT and matrices}},
\newblock \bibinfo{journal}{JHEP} \bibinfo{volume}{01} (\bibinfo{year}{2021}) \bibinfo{pages}{073}. \DOIprefix\doi{10.1007/JHEP01(2021)073}. \href{http://arxiv.org/abs/2006.07072}{{\tt arXiv:2006.07072}}.
\bibitem[{Gutperle and Strominger(2003)}]{Gutperle:2003xf}
\bibinfo{author}{M.~Gutperle}, \bibinfo{author}{A.~Strominger},
\newblock \bibinfo{title}{{Time - like boundary Liouville theory}},
\newblock \bibinfo{journal}{Phys. Rev. D} \bibinfo{volume}{67} (\bibinfo{year}{2003}) \bibinfo{pages}{126002}. \DOIprefix\doi{10.1103/PhysRevD.67.126002}. \href{http://arxiv.org/abs/hep-th/0301038}{{\tt arXiv:hep-th/0301038}}.
\bibitem[{Polyakov(1981)}]{Polyakov1981QuantumGO}
\bibinfo{author}{A.~M. Polyakov},
\newblock \bibinfo{title}{Quantum geometry of bosonic strings},
\newblock \bibinfo{journal}{Physics Letters B} \bibinfo{volume}{103} (\bibinfo{year}{1981}) \bibinfo{pages}{207--210}. \URLprefix \url{https://api.semanticscholar.org/CorpusID:121709051}.
\bibitem[{Seiberg(1990)}]{seiberg_notes}
\bibinfo{author}{N.~Seiberg},
\newblock \bibinfo{title}{{Notes on Quantum Liouville Theory and Quantum Gravity}},
\newblock \bibinfo{journal}{Progress of Theoretical Physics Supplement} \bibinfo{volume}{102} (\bibinfo{year}{1990}) \bibinfo{pages}{319--349}. \URLprefix \url{https://doi.org/10.1143/PTP.102.319}. \DOIprefix\doi{10.1143/PTP.102.319}.
\bibitem[{Suzuki and Takayanagi(2021)}]{Suzuki:2021zbe}
\bibinfo{author}{K.~Suzuki}, \bibinfo{author}{T.~Takayanagi},
\newblock \bibinfo{title}{{JT gravity limit of Liouville CFT and matrix model}},
\newblock \bibinfo{journal}{JHEP} \bibinfo{volume}{11} (\bibinfo{year}{2021}) \bibinfo{pages}{137}. \DOIprefix\doi{10.1007/JHEP11(2021)137}. \href{http://arxiv.org/abs/2108.12096}{{\tt arXiv:2108.12096}}.
\bibitem[{Collier et~al.(2023)Collier, Eberhardt, Mühlmann, and Rodriguez}]{collier2023virasoro}
\bibinfo{author}{S.~Collier}, \bibinfo{author}{L.~Eberhardt}, \bibinfo{author}{B.~Mühlmann}, \bibinfo{author}{V.~A. Rodriguez},
\newblock \bibinfo{title}{The virasoro minimal string}  (\bibinfo{year}{2023}). \href{http://arxiv.org/abs/2309.10846}{{\tt arXiv:2309.10846}}.
\bibitem[{{Guillarmou} et~al.(2023){Guillarmou}, {Kupiainen}, and {Rhodes}}]{2023_remi}
\bibinfo{author}{C.~{Guillarmou}}, \bibinfo{author}{A.~{Kupiainen}}, \bibinfo{author}{R.~{Rhodes}},
\newblock \bibinfo{title}{{Compactified Imaginary Liouville Theory}}  (\bibinfo{year}{2023}). \href{http://arxiv.org/abs/2310.18226}{{\tt arXiv:2310.18226}}.
\bibitem[{Seiberg and Shih(2005)}]{Seiberg:2004at}
\bibinfo{author}{N.~Seiberg}, \bibinfo{author}{D.~Shih},
\newblock \bibinfo{title}{{Minimal string theory}},
\newblock \bibinfo{journal}{Comptes Rendus Physique} \bibinfo{volume}{6} (\bibinfo{year}{2005}) \bibinfo{pages}{165--174}. \DOIprefix\doi{10.1016/j.crhy.2004.12.007}. \href{http://arxiv.org/abs/hep-th/0409306}{{\tt arXiv:hep-th/0409306}}.
\bibitem[{Zamolodchikov(2005)}]{Zamolodchikov:2005fy}
\bibinfo{author}{A.~B. Zamolodchikov},
\newblock \bibinfo{title}{{Three-point function in the minimal Liouville gravity}},
\newblock \bibinfo{journal}{Theor. Math. Phys.} \bibinfo{volume}{142} (\bibinfo{year}{2005}) \bibinfo{pages}{183--196}. \DOIprefix\doi{10.1007/s11232-005-0003-3}. \href{http://arxiv.org/abs/hep-th/0505063}{{\tt arXiv:hep-th/0505063}}.
\bibitem[{Kazakov(1989)}]{Kazakov:1989bc}
\bibinfo{author}{V.~A. Kazakov},
\newblock \bibinfo{title}{{The Appearance of Matter Fields from Quantum Fluctuations of 2D Gravity}},
\newblock \bibinfo{journal}{Mod. Phys. Lett. A} \bibinfo{volume}{4} (\bibinfo{year}{1989}) \bibinfo{pages}{2125}. \DOIprefix\doi{10.1142/S0217732389002392}.
\bibitem[{Johnson(2021)}]{Johnson:2020superJT}
\bibinfo{author}{C.~V. Johnson},
\newblock \bibinfo{title}{{Jackiw-Teitelboim supergravity, minimal strings, and matrix models}},
\newblock \bibinfo{journal}{Phys. Rev. D} \bibinfo{volume}{103} (\bibinfo{year}{2021}) \bibinfo{pages}{046012}. \DOIprefix\doi{10.1103/PhysRevD.103.046012}. \href{http://arxiv.org/abs/2005.01893}{{\tt arXiv:2005.01893}}.
\bibitem[{Johnson(2024)}]{Johnson:2024bue}
\bibinfo{author}{C.~V. Johnson},
\newblock \bibinfo{title}{{On the Random Matrix Model of the Virasoro Minimal String}}  (\bibinfo{year}{2024}). \href{http://arxiv.org/abs/2401.06220}{{\tt arXiv:2401.06220}}.
\bibitem[{{Kaufmann} et~al.(1996){Kaufmann}, {Manin}, and {Zagier}}]{1996CMaPh.181..763K}
\bibinfo{author}{R.~{Kaufmann}}, \bibinfo{author}{Y.~{Manin}}, \bibinfo{author}{D.~{Zagier}},
\newblock \bibinfo{title}{{Higher Weil-Petersson volumes of moduli spaces of stable n-pointed curves}},
\newblock \bibinfo{journal}{Communications in Mathematical Physics} \bibinfo{volume}{181} (\bibinfo{year}{1996}) \bibinfo{pages}{763--787}. \DOIprefix\doi{10.1007/BF02101297}. \href{http://arxiv.org/abs/alg-geom/9604001}{{\tt arXiv:alg-geom/9604001}}.
\bibitem[{Mirzakhani(2007)}]{Mirzakhani}
\bibinfo{author}{M.~Mirzakhani},
\newblock \bibinfo{title}{Weil-petersson volumes and intersection theory on the moduli space of curves},
\newblock \bibinfo{journal}{Journal of The American Mathematical Society} \bibinfo{volume}{20} (\bibinfo{year}{2007}) \bibinfo{pages}{1--23}. \DOIprefix\doi{10.1090/S0894-0347-06-00526-1}.
\bibitem[{Brezin and Kazakov(1990)}]{Brezin:1990rb}
\bibinfo{author}{E.~Brezin}, \bibinfo{author}{V.~A. Kazakov},
\newblock \bibinfo{title}{{Exactly Solvable Field Theories of Closed Strings}},
\newblock \bibinfo{journal}{Phys. Lett. B} \bibinfo{volume}{236} (\bibinfo{year}{1990}) \bibinfo{pages}{144--150}. \DOIprefix\doi{10.1016/0370-2693(90)90818-Q}.
\bibitem[{Moore et~al.(1991)Moore, Seiberg, and Staudacher}]{Moore:1991ir}
\bibinfo{author}{G.~W. Moore}, \bibinfo{author}{N.~Seiberg}, \bibinfo{author}{M.~Staudacher},
\newblock \bibinfo{title}{{From loops to states in 2-D quantum gravity}},
\newblock \bibinfo{journal}{Nucl. Phys. B} \bibinfo{volume}{362} (\bibinfo{year}{1991}) \bibinfo{pages}{665--709}. \DOIprefix\doi{10.1016/0550-3213(91)90548-C}.
\bibitem[{Belavin and Zamolodchikov(2009)}]{Belavin:2008kv}
\bibinfo{author}{A.~A. Belavin}, \bibinfo{author}{A.~B. Zamolodchikov},
\newblock \bibinfo{title}{{On Correlation Numbers in 2D Minimal Gravity and Matrix Models}},
\newblock \bibinfo{journal}{J. Phys. A} \bibinfo{volume}{42} (\bibinfo{year}{2009}) \bibinfo{pages}{304004}. \DOIprefix\doi{10.1088/1751-8113/42/30/304004}. \href{http://arxiv.org/abs/0811.0450}{{\tt arXiv:0811.0450}}.
\bibitem[{Okuyama and Sakai(2020)}]{Okuyama:2019xbv}
\bibinfo{author}{K.~Okuyama}, \bibinfo{author}{K.~Sakai},
\newblock \bibinfo{title}{{JT gravity, KdV equations and macroscopic loop operators}},
\newblock \bibinfo{journal}{JHEP} \bibinfo{volume}{01} (\bibinfo{year}{2020}) \bibinfo{pages}{156}. \DOIprefix\doi{10.1007/JHEP01(2020)156}. \href{http://arxiv.org/abs/1911.01659}{{\tt arXiv:1911.01659}}.
\bibitem[{Johnson and Rosso(2021)}]{Johnson:2020lns}
\bibinfo{author}{C.~V. Johnson}, \bibinfo{author}{F.~Rosso},
\newblock \bibinfo{title}{{Solving Puzzles in Deformed JT Gravity: Phase Transitions and Non-Perturbative Effects}},
\newblock \bibinfo{journal}{JHEP} \bibinfo{volume}{04} (\bibinfo{year}{2021}) \bibinfo{pages}{030}. \DOIprefix\doi{10.1007/JHEP04(2021)030}. \href{http://arxiv.org/abs/2011.06026}{{\tt arXiv:2011.06026}}.
\bibitem[{Castro(2023)}]{Castro:2023rfd}
\bibinfo{author}{A.~Castro},
\newblock \bibinfo{title}{{Critical JT gravity}},
\newblock \bibinfo{journal}{JHEP} \bibinfo{volume}{08} (\bibinfo{year}{2023}) \bibinfo{pages}{036}. \DOIprefix\doi{10.1007/JHEP08(2023)036}. \href{http://arxiv.org/abs/2306.14823}{{\tt arXiv:2306.14823}}.
\bibitem[{Ambjorn et~al.(1990)Ambjorn, Jurkiewicz, and Makeenko}]{Ambjorn:1990ji}
\bibinfo{author}{J.~Ambjorn}, \bibinfo{author}{J.~Jurkiewicz}, \bibinfo{author}{Y.~M. Makeenko},
\newblock \bibinfo{title}{{Multiloop correlators for two-dimensional quantum gravity}},
\newblock \bibinfo{journal}{Phys. Lett. B} \bibinfo{volume}{251} (\bibinfo{year}{1990}) \bibinfo{pages}{517--524}. \DOIprefix\doi{10.1016/0370-2693(90)90790-D}.
\bibitem[{Budd(2022)}]{Budd:2020rbm}
\bibinfo{author}{T.~Budd},
\newblock \bibinfo{title}{{Irreducible Metric Maps and Weil\textendash{}Petersson Volumes}},
\newblock \bibinfo{journal}{Commun. Math. Phys.} \bibinfo{volume}{394} (\bibinfo{year}{2022}) \bibinfo{pages}{887--917}. \DOIprefix\doi{10.1007/s00220-022-04418-6}. \href{http://arxiv.org/abs/2012.11318}{{\tt arXiv:2012.11318}}.
\bibitem[{Saad et~al.(2019)Saad, Shenker, and Stanford}]{Saad:2019lba}
\bibinfo{author}{P.~Saad}, \bibinfo{author}{S.~H. Shenker}, \bibinfo{author}{D.~Stanford},
\newblock \bibinfo{title}{{JT gravity as a matrix integral}}  (\bibinfo{year}{2019}). \href{http://arxiv.org/abs/1903.11115}{{\tt arXiv:1903.11115}}.
\bibitem[{Johnson(2021)}]{Johnson:2020exp}
\bibinfo{author}{C.~V. Johnson},
\newblock \bibinfo{title}{{Explorations of nonperturbative Jackiw-Teitelboim gravity and supergravity}},
\newblock \bibinfo{journal}{Phys. Rev. D} \bibinfo{volume}{103} (\bibinfo{year}{2021}) \bibinfo{pages}{046013}. \DOIprefix\doi{10.1103/PhysRevD.103.046013}. \href{http://arxiv.org/abs/2006.10959}{{\tt arXiv:2006.10959}}.

\end{thebibliography}

\end{document}